\documentclass[universe,article,accept,pdftex,moreauthors]{Definitions/mdpi} 
\firstpage{1} 
\pubvolume{1}
\issuenum{1}
\articlenumber{0}
\pubyear{2023}
\copyrightyear{2023}
\datereceived{ } 
\daterevised{ } 
\dateaccepted{ } 
\datepublished{ } 
\hreflink{https://doi.org/} 

\Title{Magnetized black holes: interplay between charge and rotation}

\TitleCitation{Magnetized black holes}

\Author{Vladim\'{\i}r Karas $^{1,\ddagger}$\orcidA{}, and Zden\v{e}k Stuchl\'{\i}k $^{2,\ddagger}$\orcidB{}}

\AuthorNames{Vladim\'{\i}r Karas and Zden\v{e}k Stuchl\'{\i}k}

\AuthorCitation{Karas, V.; Stuchl\'{\i}k, Z.}

\address{%
$^{1}$ \quad Astronomical~Institute, Czech Academy~of~Sciences, Bo\v{c}n\'{\i}~II~1401, CZ-14100 Prague, Czech~Republic; vladimir.karas@asu.cas.cz\\
$^{2}$ \quad Research Centre of Theoretical Physics and Astrophysics, Institute of Physics, Silesian University in Opava, Bezru\v{c}ovo n\'{a}m.\ 13, CZ-74601 Opava, Czech Republic; zdenek.stuchlik@physics.slu.cz}

\corres{Correspondence: vladimir.karas@asu.cas.cz}

\secondnote{These authors contributed equally to this work.}

\abstract{Already in the cornerstone works on astrophysical black holes published as early as in 1970s, Ruffini and collaborators have revealed potential importance of an intricate interaction between the effects of strong gravitational and electromagnetic fields. Close to the event horizon of the black hole, magnetic and electric lines of force become distorted and dragged even a in purely electro-vacuum system. Moreover, as the plasma effects inevitably arise in any astrophysically realistic environment, particles of different electric charge can separate from each other, become accelerated away from the black hole or accreted onto it, and contribute to the net electric charge of the black hole. From the point of principle, the case of super-strong magnetic fields is of particular interest, as the electromagnetic field can act as a source of gravity and influence the space-time geometry. In a brief celebratory note we revisit aspects of rotation and charge within the framework of exact (asymptotically non-flat) solutions of mutually coupled Einstein-Maxwell equations that describe magnetized, rotating black holes.}


\begin{document}

\newcommand{\beq}{\begin{equation}}
\newcommand{\eeq}{\end{equation}}
\newcommand{\rd}{{\,\rm d}}
\newcommand{\ri}{{\rm{i}}}
\newcommand{\nab}{\mbox{\protect\boldmath$\nabla$}}
\renewcommand{\vec}[1]{\mbox{\protect\boldmath$#1$}}
\newcommand{\cl}{\mathcal{L}}
\newcommand{\ce}{\mathcal{E}}
\newcommand{\dd}{\mathrm{d}}
\def\p{p} 
\def\cp{\pi} 
\def\af{\zeta} 
\def\x{x}
\def\HP{\widetilde{H_{\rm P}}}
\newcommand{\cb}{{\cal{B}}}
\newcommand{\mnras}{Mon. Not. RAS}
\newcommand{\prd}{Phys. Rev. D}
\newcommand{\jcap}{JCAP}
\newcommand{\prl}{Phys. Rev. Lett.}
\newcommand{\apjl}{Astrophys. J. Lett.}
\newcommand{\apj}{Astrophys. J.}
\newcommand{\aap}{Astron. Astrophys.}
\newcommand{\apss}{Astrophys. Space Sci.}
\def\bear{\begin{eqnarray}}
\def\ear{\end{eqnarray}}
\def\nn{\nonumber\\}
\def\Scw{Schwarz\-schild}
\section{Introduction}

Classical black holes are described by a small number of such parameters, in particular, the mass, electric and magnetic charges, and the angular momentum (spin) \citep{cha83,dew73}. As a model of cosmic black holes, these objects are spatially localized and they lack any surface; the resulting space-time has, by assumption, no material content in the form of fluids that could contribute as a source of the gravitational field. These objects do not support their own magnetic field: just gravito-magnetical component is induced by rotation \citep{pun08}. The interacting magnetic field to which astrophysical black holes are embedded is of external origin (Ruffini \& Wilson \citep{ruf75}), although it may naturally interact with the Kerr-Newman intrinsic charge \cite{new65}. 

This approach was employed by a number of authors to address the problem of electromagnetic effects near a rotating (Kerr) black hole. On the other hand, self-consistent solutions of coupled Einstein-Maxwell equations for black holes immersed in electromagnetic fields have been studied only within stationary, axially symmetric, electro-vacuum models. It has soon appeared that the test electromagnetic field approximation was fully adequate for modelling astrophysical sources, however, the long-term evolution of magnetospheres of rotating black holes and the consequences of strong gravity remained still open to further work \cite{bae86,ern76}. To explore the latter, intriguing effects of ultra-strong magnetic fields, we employ an axially symmetric solution that was derived originally in 1970s in terms of magnetization techniques \citep{ern76,kra80}. 

While the main aim and the motivation of our present contribution is to briefly summarize some of the aspects of magnetized black holes that have been explored over six decades of intensive research, and where the honoree and his collaborators published a number of widely cited discoveries, we will mention also some interesting features of the induced electric charge that occur in this regime and are explored to date. In fact, the generation of magnetic fields goes hand in hand with creation of corresponding electric fields which always arise in moving media and, for that matter, they appear once a rotating body is involved.

\section{Magnetized Kerr-Newman black hole in charge equlibrium}
We can write the system of mutually coupled, Einstein-Maxwell equations (Chandrasekhar 1983 \citep{cha83}),
\beq 
R_{\mu\nu}-\textstyle{\frac{1}{2}}Rg_{\mu\nu}=8\pi T_{\mu\nu},
\eeq
where the source term $T_{\mu\nu}$ is of purely electromagnetic origin,
\beq
T^{\alpha\beta}\equiv T^{\alpha\beta}_{\rm EMG}=\frac{1}{4\pi}\left(F^{\alpha\mu}F^\beta_\mu- \frac{1}{4}F^{\mu\nu}F_{\mu\nu}g^{\alpha\beta}\right),
\eeq
and $^\star F_{\mu\nu}\equiv\frac{1}{2}{\varepsilon_{\mu\nu}}^{\rho\sigma}F_{\rho\sigma}$.
Let us first consider a strongly magnetized Kerr-Newman (MKN) black hole. This is an electro-vacuum space-time solution with a regular event horizon that satisfies the conditions of axial symmetry and stationarity. Hence, it adopts a general form \citep{rom14,wal84}
\beq
\rd s^2=f^{-1}\left[e^{2\gamma}\left(\rd z^2+\rd\rho^2\right)+\rho^2\rd\phi^2\right]-f\left(\rd t-\omega\rd\phi\right)^2,
\eeq
with $f$, $\omega$, and $\gamma$ being the functions of cylindrical coordinates $\rho$ and $z$ only because of the assumed symmetries. Although in the weak electromagnetic field approximation the Kerr metric gives the line element \cite{ker63}, the case of strong magnetic field is different especially at large values of the cylindrical radius. This is because of magnetic field curving the spacetime and changing its asymptotical characteristics into a non-flat (cosmological) solution (see, e.g., Gal'tsov 1986 \citep{gal86}). 


Christodoulou  and Ruffini \citep{chr73} introduced the magnetic and electric lines of force that are defined, respectively, by the direction of Lorentz force that acts on electric/magnetic charges,
\beq
\frac{\rd u^\mu}{\rd\tau}\propto \,^{\star}\!F^\mu_\nu\,u^\nu,\qquad \frac{\rd u^\mu}{\rd\tau}\propto F^\mu_\nu\,u^\nu.
\eeq
In an axially symmetric system, the equation for magnetic lines of force adopts an a form that is fully expected on the basis of classical electromagnetism,
\beq
\frac{\rd r}{\rd\theta}=-\frac{F_{\theta\phi}}{F_{r\phi}},\qquad \frac{\rd r}{\rd\phi}=\frac{F_{\theta\phi}}{F_{r\theta}}.
\eeq
%

By employing the solution generating technique \citep{kin73}, Garc\'{\i}a D\'{\i}az 1985 \citep{gar85} gave a very general and explicit form of the {\em exact} spacetime metric of a strongly magnetized black hole:
\beq
ds^2=|\Lambda|^2\Sigma\left(\Delta^{-1}\rd{r}^2+\rd{\theta}^2-\Delta{A^{-1}}\rd{t}^2\right) 
+|\Lambda|^{-2}\Sigma^{-1}A\sin^2\theta\left(\rd{\phi}-\omega\rd{t}\right)^2,
\label{g}
\eeq
where $\Sigma(r,\theta)=r^2+a^2\cos^2\theta$, $\Delta(r)=r^2-2Mr+a^2+e^2$, $A(r,\theta)=(r^2+a^2)^2-{\Delta}a^2\sin^2\theta$ are the well-known metric functions from the Kerr-Newman solution. The event horizon exists for $a^2+e^2\leq1$. In the magnetized case, because of asymptotically non-flat nature of the spacetime, the parameters $a$ and $e$ are not identical with the black hole total spin and electric charge \citep{his81}. Moreover, because of asymptotically non-flat nature of the spacetime, the Komar-type angular momentum and electric charge (as well as the black hole mass) have to be defined by integration over the horizon sphere rather than at radial infinity \citep{kar90}.
The magnetization function $\Lambda=1+\beta\Phi-\frac{1}{4}\beta^2\mathcal{E}$ is given in terms of the Ernst potentials $\Phi(r,\theta)$ and $\mathcal{E}(r,\theta)$,
\begin{eqnarray}
\Sigma\Phi
 &=& ear\sin^2\theta-{\Im}e\left(r^2+a^2\right)\cos\theta, \\
\Sigma\mathcal{E}
 &=& -A\sin^2\theta-e^2\left(a^2+r^2\cos^2\theta\right)
 \nonumber \nonumber \\
 & & + 2{\Im}a\left[\Sigma\left(3-\cos^2\theta\right)+a^2\sin^4\theta-
 re^2\sin^2\theta\right]\cos\theta.
\end{eqnarray}

\begin{figure}[tbh!]
\begin{center}
\includegraphics[width=0.9\textwidth]{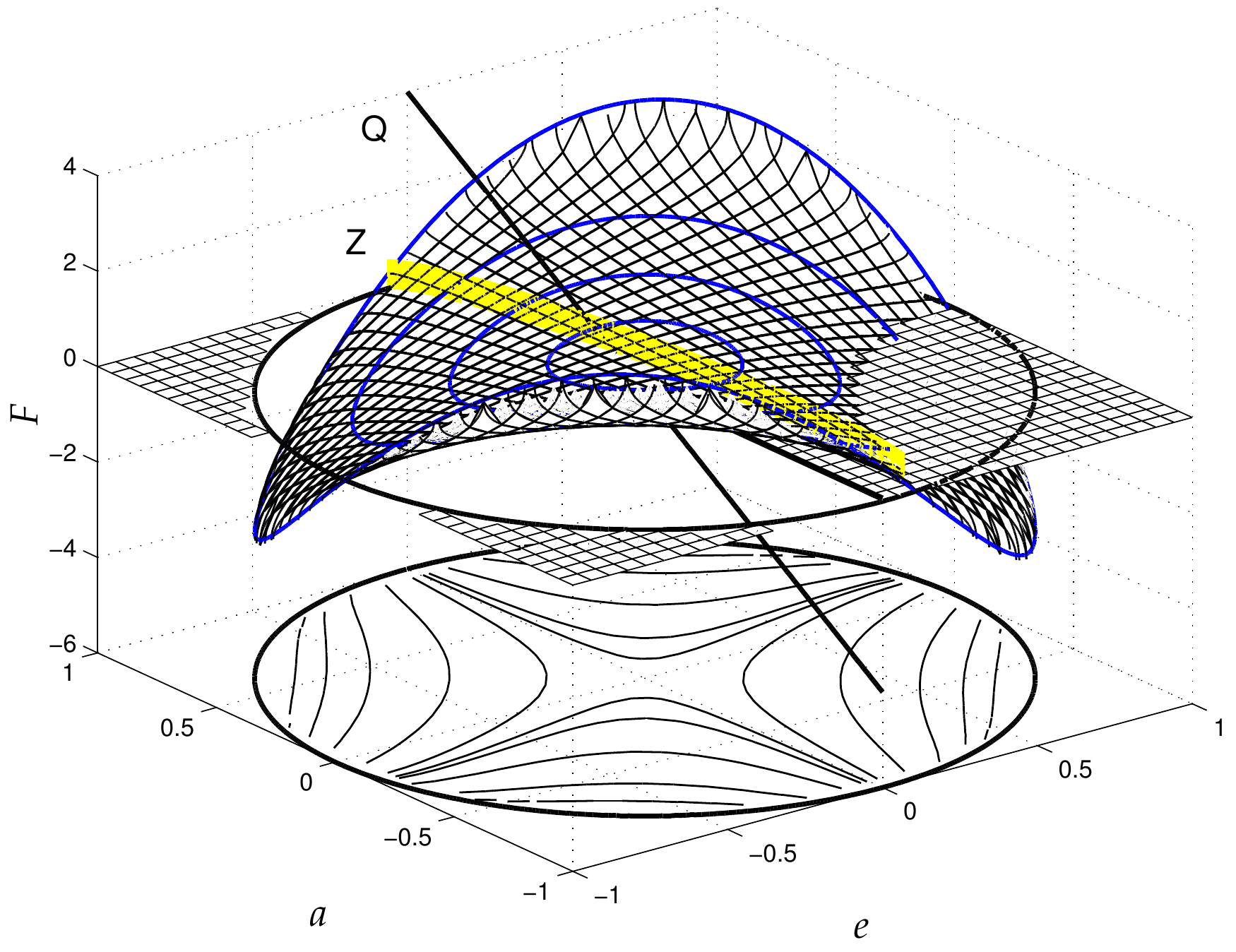}
\end{center}
\caption{Surface plot of the magnetic flux function, $F(a,e)$, across a hemisphere bounded by $\theta=\pi/2$ and located on the MKN black hole horizon. A fixed value of the magnetization parameter $\beta=0.05$ has been selected. Projected contours are also shown for an improved clarity of the plot. The surface is restricted by the condition for the emergence of event horizon, $a^2+e^2\leq1$. Four circles of $\sqrt(a^2+e^2)=0.25$, 0.5, 0.75, and 1.0 are shown to guide eye. The yellow band on the surface, denoted by ``{\sf{Z}}'', indicates where the total electric charge is zero. Note: Unlike the case of weakly magnetized black hole, the moment of vanishing charge does {\em not} coincide with zero of the charge parameter, $e=0$. On the other hand, $Q(a,e=0)$ does not vanish and its graph is shown by solid curve ``{\sf{Q}}''. This is the feature of exact MKN metric, where the two nulls do not generally coincide, as further detailed in \cite{kar90} (this figure has been reproduced with persion from \textit{Physica Scripta} article ref.\ \citep{kar00}).}
\label{fig1}
\end{figure}

The  components of the electromagnetic field with respect to orthonormal LNRF components are
\begin{eqnarray}
H_{(r)}+{\ri}E_{(r)} &=& A^{-1/2}\sin^{-1}\!\theta\,\Phi^{\prime}_{,\theta},\\
H_{(\theta)}+{\ri}E_{(\theta)} &=&-\left(\Delta/A\right)^{1/2}\sin^{-1}\!\theta\,\Phi^{\prime}_{,r},
\label{heth}
\end{eqnarray}
where $\Phi^{\prime}(r,\theta)=\Lambda^{-1}\left(\Phi-\frac{1}{2}\beta\mathcal{E}\right)$,
and the total electric charge $Q_{\rm{H}}$ is
\begin{equation}
Q_{\rm{H}} = -|\Lambda_0|^2\,\Im{\rm{m}\,}\Phi^{\prime}\left(r_+,0\right).
\label{qh} 
\end{equation}
The magnetic flux $\Phi_{\rm{m}}(\theta)$ across a cap placed in an axisymmetric position on the horizon is then \citep{kar00}
\begin{equation}
\Phi_{\rm{m}} = 2\pi|\Lambda_0|^2\,\Re{\rm{e}\,}\Phi^{\prime}
 \left(r_+,\bar{\theta}\right)\Bigr|\strut^{\theta}_{\bar{\theta}=0},
\label{fh}
\end{equation}
where $\Lambda_0=\Lambda(\theta=0)$. In Figure \ref{fig1}, the surface plot of the magnetic flux $F$ across the hemisphere $\theta=\pi/2$ is shown as a function of spin parameter $a$ and the electric charge parameter $e$. The surface on the horizon is defined on the circle $a^2+e^2\leq1$. 

The definition interval of the azimuthal coordinate in the magnetized solution needs to be rescaled by a factor $\Lambda_0$ (not to be confused with cosmological term) in order to avoid a conical singularity on the symmetry axis \cite{his81}, which effectively leads to the increase of the horizon surface area, and thereby also the total magnetic flux threading the event horizon \cite{kar88}. Let us note that cosmic magnetic fields are limited in strength only by quantum theory effects. In highly magnetized rotators the energy of the magnetic field can be converted into high-energy gamma rays, but such mechanisms require over $10^{12}$ tesla; we shall not consider this ultra-strong magnetic fields in the rest of the paper.

The above discussed electro-vacum solutions need to be extended by including  an electrically conducting plasma. Once this is introduced into the MKN system, one needs to clarify to what extent the newly emerging role of the $\Lambda$ term affects the characteristics of the flow of material. This can be investigated in terms of {\em plasma horizon} and the {\em guiding centre} approximation, which was originally introduced in the context of accreting black holes by Ruffini \cite{ruf79}, Damour et al. \cite{dam78}, and Hanni \& Valdarnini \cite{han79}. 
Surfaces of magnetic support were further extended to the case of a black hole that is moving at constant velocity \cite{lyu11,mor14}. Although these authors considered the case of weak (test) magnetic field in Kerr metric, in a subsequent analysis by Karas \& Vokrouhlick\'y \citep{kar91} we verified that, for astrophysically realistic values of magnetic intensity, the approximate flow lines coincide almost precisely with those constructed for the exact MKN system; they are indistinguishable for practical purposes.  

The energy density contained in astrophysically realistic electromagnetic fields turns out to be far too low to influence the space–time noticeably. Test-field solutions are thus adequate for describing weak electromagnetic fields, even those around magnetized neutron stars and cosmic black holes that are currently known.

\section{Weak magnetic field and particle acceleration}
For strong influence of the external magnetic field on the spacetime structure of the black hole, its intensity has to be enormously high, comparable with 
\beq
               B_{\rm GR} = 10^{18}\; \frac{10M_{\odot}}{M} \; [{\rm G}] .  
\eeq 
Realistic magnetic fields in astrophysical situations are strongly under this limit, even in the case of fields near magnetars, reaching $B \sim 10^{15}$~gauss. Therefore, for the astrophysical processes we can usually put the magnetic spacetime factor $\Lambda=1$ and the electric charge $e=0$, using the canonical, asymptotically flat Kerr metric. As for the electromagnetic term, an asymptotically uniform magnetic field, orthogonal to the spacetime equatorial plane, can then be determined by the electromagnetic 4-vector potential taking the form  
\beq 
A_{t} = \frac{B}{2} (g_{t\phi} + 2 a g_{tt}) - \frac{Q}{2} g_{tt} - \frac{Q}{2}, \quad
A_{\phi} = \frac{B}{2} (g_{\phi\phi} + 2 a g_{t\phi}) - \frac{Q}{2} g_{t\phi}, 
\eeq
where the induced electric charge of the black hole $Q$ is also introduced. For non-charged black holes there is $Q=0$, the maximal induced black hole charge generated by the black hole rotation takes the Wald value $Q_{W}=2aB$ (or $Q_{W}=2aBM$ if we keep the mass term) -- see \cite{wal84}; the influence of the induced so called Wald charge on the spacetime structure could be also abandoned \cite{bic80,kin75}. For black holes with the maximal Wald charge we arrive to the electromagnetic potential  
\beq 
A_{t} = \frac{B}{2} g_{t\phi} - \frac{Q_{W}}{2}, \quad
A_{\phi} = \frac{B}{2} g_{\phi\phi} . 
\eeq
It is crucial that even in this case the $A_{t}$ component remains non-zero and can lead to very strong acceleration mechanism for sufficiently massive black holes and strong magnetic fields \cite{Stu-etal:2020:Universe:}. The significant role of the electromagnetic fields in processes near a black hole horizon was for the first time presented in a series of works of Ruffini and his collaborators in \cite{Ruf:1973:BlackHoles:}. It could be well demonstrated for the charged test particle motion in the case of ionized Keplerian disks \cite{Stu-etal:2020:Universe:}.

Motion of an electrically charged test particle with charge $q$ and mass $m$ is determined by the Lorentz equation
\beq
           m\frac{Du^{\mu}}{D\tau} = qF^{\mu}_{\nu}u^{\nu} ,
\eeq 
where $\tau$ is the particle proper time, and $F^{\mu}_{\nu}$ is the Faraday tensor of the electromagnetic field. For the Kerr-Newman black holes the Lorentz equations can be separated and given in terms of first integrals, governing thus fully regular test particle motion \cite{Car:1973:BlaHol:,cha83,wal72}, while for magnetized Kerr black holes the separability is impossible implying generally chaotic character of the motion \cite{kar92,luk14,pan19,Stu-etal:2020:Universe:}.

Nevertheless, due to the symmetries of the magnetized Kerr black holes with the uniform magnetic field lines orthogonal to the equatorial plane of the spacetime, we can introduce Hamiltonian in the form 
\beq
  H =  \textstyle{\frac{1}{2}} g^{\alpha\beta} (\cp_\alpha - q A_\alpha)(\cp_\beta - q A_\beta) + \textstyle{\frac{1}{2}} \, m^2, 
\eeq
where the canonical four-momentum $\cp^\mu = \p^\mu + q A^\mu$ is related to the kinematic four-momentum $\p^\mu = m u^\mu$ and the influence of the electromagnetic field reflected by $q A^\mu$. The motion is then governed by the Hamilton equations
\beq
 \frac{\dd \x^\mu}{\dd \af} \equiv \p^\mu = \frac{\partial H}{\partial \cp_\mu}, \quad
 \frac{\dd \cp_\mu}{\dd \af} = - \frac{\partial H}{\partial \x^\mu} ; 
\eeq
the affine parameter is related to the particle proper time as $\af=\tau/m$. 

Due to the background symmetries we can introduce two constants of the motion: energy $E$ and angular momentum $L$ as conserved components of the canonical momentum read
\beq
-E = \pi_t = g_{tt} p^t + g_{t\phi} p^{\phi} + q A_{t},
\eeq 
\beq
L = \pi_\phi = g_{\phi\phi} p^\phi + g_{\phi t} p^{t} + q A_{\phi}. 
\eeq
Introducing the specific energy $\ce=E/m$, the specific axial angular momentum $\cl=L/m$, and the magnetic interaction parameter $\cb = q B / 2m$, we obtain Hamiltonian with two degrees of freedom, and the four-dimensional phase space $\{r,\theta;\p_r,\p_\theta\}$ in the form 
\beq 
       H = \textstyle{\frac{1}{2}} g^{rr} \, \p_r^2 + \textstyle{\frac{1}{2}} g^{\theta\theta} \, \p_\theta^2 + \HP(r,\theta) , 
\eeq
enabling introduction of the effective potential of the radial and latitudinal motion. The energy condition relates the specific energy to the effective potential as  
\beq
        \ce = V_{\rm eff}(r,\theta) 
\eeq
where				
\beq
       V_{\rm eff} (r,\theta) = \frac{ -\beta + \sqrt{\beta^2 - 4\alpha\gamma} }{ 2 \alpha }, 
\eeq
with
\beq
\beta  = 2 [ g^{t\phi} (\cl-\tilde{q}A_{\phi}) - g^{tt}\tilde{q}A_{t} ], \qquad \alpha =  -g^{tt},
\eeq
and
\beq
\gamma  =  -g^{\phi\phi}(\cl-\tilde{q}A_{\phi})^2 - g^{tt} \tilde{q}^2 A_{t}^2 + 2 g^{t\phi}\tilde{q} A_{t} (\cl-\tilde{q}A_{\phi}) -1.
\eeq
The effective potential defined here is properly chosen for the region above the outer horizon of the black hole, governing the regions allowed for the motion of a charged particle with fixed value of the axial angular momentum. 

Study of the motion of charged particles applied to the case of ionized Keplerian disks (see \cite{Stu-etal:2020:Universe:} for a review) demonstrates that the fate of the ionized disks depends on the magnetic interaction parameter. In the so called gravitational regime when gravity is suppressing the role of the electromagnetic field ($\cb \ll 1$), motion of the particles of the ionized Keplerian disks can be considered as being in quasi-circular harmonic epicyclic motion of regular character, enabling explanation of high-frequency quasi-periodic oscillations of X-rays observed in microquasars and some active galactic nuclei \cite{Stu-Kol-Tur:2022:PASJ:}. In the so called gravity-magnetic regime when the role of both fields is comparable ($\cb \sim 1$), the motion is fully chaotic, leading generally to toroidal configurations. In the so called magnetic regime ($\cb \gg 1$), the role of the magnetic field is decisive, and the motion could have finally a regular character governed by the Larmor precession frequency. 

In the case of $\cb > 1$ a special effect of chaotic scattering can be relevant \cite{kop10,Stu-Kol:2016:EPJC:} when the ionized particle can be accelerated along the magnetic field lines after a period of chaotic motion that decreases with increasing magnetic parameter \cite{kar21}. In such situations the magnetic Penrose process could be realized with extremely high efficiency.  The tentative magnetic Penrose process (MPP; see \cite{Stu-Kol-Tur:2021:Universe:}) is a local decay processes; its energy balance is governed by the local value of the electromagnetic field (potential) -- for this reason the simple approximation of asymptotically uniform magnetic field aligned with the rotations axis can be well applied \cite{Stu-etal:2020:Universe:}.

Let us consider splitting of the 1st particle with energy $E_1$ (electrically neutral or positively charged with charge $q_1$) onto two charged particles, 2nd one having a positive charge $q_2$ and the 3rd one having a negative charge $q_3$. If one of the particles (say the 3rd one) has a negative canonical energy $E_3 < 0$, then the second one should have the canonical energy $E_2 > E_1$ due to an extraction of the black hole energy because of the capture of the 3rd particle. The process of the split of the 1st particle into the 2nd and 3rd ones is governed by the conservation laws \cite{Stu-Kol-Tur:2021:Universe:}.  

The efficiency of the MPP is defined by relating the gained and input energies 
\beq
   \eta = \frac{E_2 - E_1}{E_1} = \frac{-E_3}{E_1} ,  
\eeq
implying the relation \cite{Tur-etal:2020:ApJ:}
\beq
   \eta_{\rm MPP} = \chi - 1 + \frac{\chi q_1 A_t - q_2A_t}{E_1} . 
\eeq
The MPP demonstrates three substantially different efficiency regimes. The low efficiency regime corresponds to the original Penrose process involving only electrically neutral particles (or vanishing electromagnetic field) with efficiency \cite{Pen:1969:NuoCim:}
\beq 
    \eta_{\rm PP(max)} = \frac{\sqrt2 - 1}{2} \sim 0.207 . 
\eeq
The moderate regime of the MPP corresponds to the situation when the electromagnetic forces are dominant, and the particles are charged, i.e., the condition $|\frac{q}{m}A_t| \gg |u_t|=|p_t|/m$ is satisfied, with efficiency approximately determined as  
\beq
       \eta^{mod}_{MPP} \sim \frac{q_2}{q1} - 1 ,
\eeq			
operating while $q_2 > q_1$. In this case the gravitationally induced electric field of the black hole is neutralized and the moderate regime of the MPP is close to the Blandford-Znajek process \cite{Bla-Zna:1977:MNRAS:}; both processes are driven by the quadrupole electric field generated due to twisting the magnetic field lines because of the spacetime frame dragging, and restricted by global neutrality of the plasma surrounding the black hole \cite{Dad-etal:2018:MNRAS:,pun08}. 
The extremely efficient regime corresponds to ionization of neutral matter and its efficiency is dominated by the term 
\beq
  \eta^{\rm extr}_{\rm MPP} \sim \frac{q_2}{m_1}A_t .
\eeq	
In the extreme regime of the MPP an enormous increase of the efficiency is possible, giving enormous energy to escaping particles. The efficiency can be as large as $\eta^{\rm extr}_{\rm MPP}\sim~10^{10}$ if the magnetic field is sufficiently large and the rotating black hole is supermassive \cite{Tur-etal:2020:ApJ:}. 

Let us note that the mechanism of charging of a boosted black hole in translatory motion has been revisited very recently, \citep{ada23,dai19}; it has attracted a renewed widespread attention because of its tentative relevance for late stages of black hole -- neutron star inspirals and their subsequent mergers. In this context, there is an interesting parallel between the {\em effects of rotation vs.\ boost}. Along a different line of research, Okamoto and Song \cite{oka23} argue that the electromagnetic self-extraction of energy will be possible only via the frame-dragged rotating magnetosphere. It will be interesting to see if the above-discussed ideas of {\em magnetic Penrose process}, where the energy extraction is explored from another view angle, will be confirmed with more accurate and complete description in the future. It seems to be very exciting that the present-day understanding is still incomplete and even controversial as the adopted approximations are tentative and await further verification or disproval \cite{san23}.


\section{Conclusions}
The MPP enables acceleration of protons and light ions up to the energy $E \sim 10^{22} eV$, corresponding to the highest-energy ultra-high energy cosmic rays (UHECR) observed on the Earth, that can occur around supermassive black holes in the active galactic nuclei similar to those in the M87 large elliptical galaxy \cite{Tur-etal:2020:ApJ:}. For accelerated electrons the energy could be even higher, but contrary to the case of protons and ions, where the back-reaction related to the synchrotron radiation of the accelerated particles is negligible, for electrons the back-reaction is extremely strong, decelerating substantially this kind of light particles -- they thus cannot be observed as UHECR \cite{Stu-Kol-Tur:2021:Universe:}. 

Our scenario is complementary to highly dynamical situations discussed in a series of articles by Ruffini et al.\ \citep{ruf08}, who explore the early, prompt phase of gamma-ray burst sources within a scenario of a baryonic shell interacting with an inhomogeneous medium (see also further references in \cite{fry14,met22,rue12,kom22,ruf18}). Whereas we do not consider temporal effects on the black hole gravitational field, we do take into account the role of magnetic field in shaping the stationary background. It turns out that for astrophysically realistic models, time dependence may be crucial. On the other hand, the impact that super-strong magnetic fields may have on the spacetime curvature is relevant with respect to our understanding of exact solutions of Einstein-Maxwell fields; this can be best revealed by employing simplified equilibrium models like the one discussed in our research note.

As a final remark, let us note that the similarity between the problem of a rotating magnetized body treated in the framework of classical electrodynamics and the corresponding black-hole electrodynamics has been widely explored in the literature (e.g. \citep{ruf73,tur20}, and numerous subsequent papers). The black hole problem seems to be more complex because we have to consider the effects of general relativity, however, the adopted space–time represents an electro-vacuum solution and it is thus idealized with a small number of free parameters. Intricate relations and numerical analysis are needed in order to determine material properties if plasma is present.

\authorcontributions{Conceptualization, V.\ Karas; Methodology, Z.\ Stuchl\'{\i}k; Writing – original draft, V.\ Karas; Writing – review \& editing, Z.\ Stuchl\'{\i}k.}

\funding{The authors thank the anonymous referees for their comments and suggestions that helped us to improve the paper and clarify several arguments. The authors acknowledge the institutional support from the Astronomical Institute in Prague and the Institute of Physics in Opava. VK acknowledges continued support from the Research Infrastructure CTA-CZ (LM2023047) funded by the Czech Ministry of Education, Youth and Sports, and the EXPRO (21-06825X) project of the Czech Science Foundation.}

\dataavailability{No new data were created or analyzed in this study. Data sharing is not applicable to this article.} 

\acknowledgments{The authors are grateful to the organizers of the {\em Conference in Celebration of Prof. Remo Ruffini 80th birthday}, held in the ICRANet seat at Villa Ratti, Nice (France) and online 16--18 May 2022, where the contribution was presented. It has been great pleasure to meet and discuss with professor Ruffini during various occasions over the years in Rome, Pescara, and elsewhere, in particular at the {\em Marcel Grossmann Meetings on General Relativity}. More recently, RR and several colleagues of ICRANet have participated at the Relativistic Astrophysics Group Workshops {\em On Black Holes and Neutron Stars} (RAGtime) that are held annually at the Institute of Physics of the Silesian University in Opava, and at the {\em 31$^{st}$ Texas Symposium on Relativistic Astrophysics} in Prague (Czech Republic).}

\conflictsofinterest{The authors declare no conflict of interest.} 

\begin{adjustwidth}{-\extralength}{0cm}

\reftitle{References}



\begin{thebibliography}{99}

\bibitem{ada23}
Adari P., Berens R., Levin J. (2023), ``Charging up boosted black holes'', \textit{Physical Review D}, 107, id.\ 044055 (15 pp.)

\bibitem{bae86}
Baez N. B., Garc\'{\i}a D\'{\i}az A. G. (1986), ``The most general magnetized Kerr-Newman metric'', \textit{J. Math. Phys.}, 27, 562

\bibitem{bic80}
Bi\v{c}\'ak~J., Dvo\v{r}\'ak~L. (1980), ``Stationary electromagnetic fields around black holes. III.'', \textit{Physical Review D}, 22, 2933


\bibitem{Bla-Zna:1977:MNRAS:}
Blandford R. D.,  Znajek R. L. (1977), ``Electromagnetic extraction of energy from Kerr black holes'', \textit{Monthly Notices of the Royal Astronomical Society}, 179, 433

\bibitem{Car:1973:BlaHol:}
Carter B. (1973), ``Black hole equilibrium states'', in Black holes (Les astres occlus), Lectures delivered at the Summer School of Theoretical Physics of the University of Grenoble at Les Houches, eds.\ C.\ DeWitt and B.\ S.\ DeWitt (New York, NY: Gordon and Breach), pp. 57--214

\bibitem{cha83}
Chandrasekhar S. (1983), The Mathematical Theory of Black Holes (Oxford: Oxford University Press)

\bibitem{chr73}
Christodoulou D., Ruffini R. (1973), ``On the electrodynamics of collapsed objects'', in Black Holes, eds. C.~DeWitt \& B.~S. DeWitt (New York: Gordon and Breach Science Publishers), p.~R151

\bibitem{Dad-etal:2018:MNRAS:}
Dadhich N.,  Tursunov A.,  Ahmedov B.,  Stuchl\'{\i}k Z. (2018), ``The distinguishing signature of magnetic Penrose process'', \textit{Monthly Notices of the Royal Astronomical Society: Letters}, 478, L89

\bibitem{dai19}
Dai Z.~G. (2019), ``Inspiral of a spinning black hole-magnetized neutron star binary: increasing charge and electromagnetic emission'', \textit{The Astrophysical Journal Letters}, 873, id.\ L13 (5 pp.)

\bibitem{dam78}
Damour T., Hanni R. S., Ruffini R., Wilson J. R. (1978), ``Regions of magnetic support of a plasma around a black hole", \textit{Physical Review D}, 17, 1518

\bibitem{dew73}DeWitt C., DeWitt B. S. (1973), Black holes. Lectures delivered at the Summer School of Theoretical Physics of the University of Grenoble at Les Houches (New York: Gordon and Breach)

\bibitem{ern76}
Ernst F.~J., and Wild W.~J. (1976), ``Kerr black holes in a magnetic universe'', \textit{J. Math.\ Phys.}, 12, 1845

\bibitem{fry14}
Fryer C. L., Rueda J. A., Ruffini R. (2014), ``Hypercritical accretion, induced gravitational collapse, and binary-driven hypernovae'', \textit {The Astrophysical Journal Letters}, 793, L36

\bibitem{gal86}  
Gal'tsov D.~V. (1986), Particles and Fields around Black Holes (Moscow: Moscow University Press)

\bibitem{gar85}
Garc\'{\i}a D\'{\i}az A. (1985), ``Magnetic generalization of the Kerr-Newman metric'', \textit{J. Math. Phys.}, 26, 155

\bibitem{han79}
Hanni R.~S., Valdarnini R. (1979), ``Magnetic support near a charged rotating black hole'', \textit{Physics Letters A}, 70, 92

\bibitem{his81}
Hiscock W.~A. (1981), ``On black holes in magnetic universes'', \textit{J.\ Math.\ Phys.}, 22, 1828

\bibitem{kar88}
Karas~V. (1988), ``Magnetic fluxes across black holes. Exact models'', \textit{Bulletin of the Astronomical Institute of Czechoslovakia}, 39, 30

\bibitem{kar00}
Karas~V., Bud\'{\i}nov\'a Z. (2000), ``Magnetic fluxes across black holes in a strong magnetic field regime'', \textit{Physica Scripta}, 61, 253

\bibitem{kar21}
Karas V., Kop\'a\v{c}ek O. (2021), ``Near horizon structure of escape zones of electrically charged particles around weakly magnetized rotating black hole: Case of oblique magnetosphere'', \textit{Astronomische Nachrichten}, 342, 357

\bibitem{kar90}
Karas~V., Vokrouhlick\'y~D. (1990), ``On interpretation of the magnetized Kerr-Newman black hole'', \textit{J.\ Math.\ Phys.}, 32, 714

\bibitem{kar91}
Karas~V., Vokrouhlick\'y~D. (1991), ``Dynamics of charged particles near a black hole in a magnetic field'', \textit{J.\ Phys.\ I France}, 1, 1005

\bibitem{kar92}
Karas V., Vokrouhlick\'y D. (1992), ``Chaotic motion of test particles in the Ernst space-time'', \textit{General Relativity and Gravitation}, 24, 729-743,

\bibitem{ker63}
Kerr, R. P. (1963), ``Gravitational field of a spinning mass as an example of algebraically special metrics'', \textit{Physical Review Letters}, 11, 237

\bibitem{kin73}
Kinnersley W. (1973), ``Generation of stationary Einstein-Maxwell fields'', \textit{J. Math. Phys.}, 14, 651

\bibitem{kin75}
King A.~R., Lasota J.~P., Kundt W. (1975), ``Black holes and magnetic fields'', \textit{Physical Review D}, 12, 3037

\bibitem{kom22}
Komissarov S. S. (2022), ``Electrically charged black holes and the Blandford-Znajek mechanism'', \textit{Monthly Notices of the Royal Astronomical Society}, 512, 2798

\bibitem{kop10}
Kop\'a\v{c}ek O., Karas V., Kov\'a\v{r} J., Stuchl\'{\i}k Z. (2010), ``Transition from regular to chaotic circulation in magnetized coronae near compact objects'', \textit{The Astrophysical Journal}, 722, 1240

\bibitem{kra80}
Kramer D., Stephani H., MacCallum M., and Herlt E. (1980), Exact Solutions of the Einstein's Field Equations (Berlin: Deutscher Verlag der Wissenschaften)

\bibitem{luk14}
Lukes-Gerakopoulos G. (2014), ``Adjusting chaotic indicators to curved spacetimes'', \textit{Physical Review D}, 89, id.\ 043002

\bibitem{lyu11}
Lyutikov M. (2011), ``Schwarzschild black holes as unipolar inductors: Expected electromagnetic power of a merger'', \textit{Physical Review D}, 83, id.\ 064001

\bibitem{met22}
Metzger, B. D. (2022), ``Luminous fast blue optical transients and type Ibn/Icn SNe from Wolf-Rayet/Black hole mergers'', \textit{The Astrophysical Journal}, 932, 84

\bibitem{mor14}
Morozova V. S., Rezzolla L., Ahmedov B. J. (2014), ``Nonsingular electrodynamics of a rotating black hole moving in an asymptotically uniform magnetic test field'', \textit{Physical Review D}, 89, id.\ 104030

\bibitem{new65}
Newman E.~T., Couch E., Chinnapared K., et al. (1965), ``Metric of a rotating, charged mass'', \textit{J. Math. Phys.}, 6, 918

\bibitem{oka23}
Okamoto I., Song Y. (2023), ``Energy self-extraction of a Kerr black hole through its frame-dragged force-free magnetosphere'', \textit{submitted}, arXiv:1904.11978v7

\bibitem{pan19}
P{\'a}nis R., Kolo{\v{s}} M, Stuchl{\'\i}k Z. (2019), ``Determination of chaotic behaviour in time series generated by charged particle motion around magnetized Schwarzschild black holes'',
 \textit{European Physical Journal C}, 79, id.\ 479,

\bibitem{Pen:1969:NuoCim:}
Penrose R. (1969), ``Gravitational collapse: the role of General Relativity'', \textit{Rivista del Nuovo Cimento}, Numero Speziale I, 252

\bibitem{pun08}
Punsly B. (2008), Black Hole Gravito-hydromagnetics (Berlin: Springer-Verlag)

\bibitem{rom14}
Romero G. E., Vila G. S. (2014), Introduction to Black Hole Astrophysics, \textit{Lecture Notes in Physics}, vol. 876 (Berlin: Springer)

\bibitem{rue12}
Rueda J. A., Ruffini R. (2012), ``On the induced gravitational collapse of a neutron star to a black hole by a type Ib/c supernova'', \textit{The Astrophysical Journal Letters}, 758, L7

\bibitem{Ruf:1973:BlackHoles:}
Ruffini R. (1973), ``On the energetics of black holes'', in Black holes (Les astres occlus), Lectures delivered at the Summer School of Theoretical Physics of the University of Grenoble at Les Houches, eds.\ C.\ DeWitt and B.\ S.\ DeWitt (New York, NY: Gordon and Breach), pp. 451--546

\bibitem{ruf79}
Ruffini R. (1979), ``On gravitationally collapsed objects'', in Relativity, Quanta and Cosmology in the Development of the Scientific Thought of Albert Einstein, eds.  M.\ Pantaleo and F.\ de Finis (Johnson Reprint Corp), pp. 599--658

\bibitem{ruf08}
Ruffini R.,  Bernardini M. G., Bianco C. L., et al. (2008), ``On gamma-ray bursts'', in The Eleventh Marcel Grossmann Meeting on Recent Developments in Theoretical and Experimental General Relativity, Gravitation and Relativistic Field Theories (Berlin, Germany, 23--29 July 2006), eds. H. Kleinert and R. T. Jantzen, ser. ed. R.~Ruffini (Singapore: World Scientific Publishing Co.)

\bibitem{ruf73}
Ruffini R., Treves A. (1973), ``On a magnetized rotating sphere'', \textit{Astrophysical Letters}, 13, 109

\bibitem{ruf18}
Ruffini R., Wang Y., Aimuratov Y., et al. (2018), ``Early X-ray flares in GRBs'', \textit{The Astrophysical Journal}, 852, id.\ 53 (27 pp.)

\bibitem{ruf75}
Ruffini R., Wilson J. R. (1975), ``Relativistic magnetohydrodynamical effects of plasma accreting into a black hole'', \textit{Physical Review D}, 12, 2959

\bibitem{san23}
Santos J., Cardoso V., Nat\'ario J. (2023), ``Electromagnetic radiation reaction and energy extraction from black holes: The tail term cannot be ignored'', \textit{Physical Review D}, id.\ 064046 (8 pp.)

\bibitem{Stu-Kol:2016:EPJC:}
Stuchl\'{\i}k Z.,  Kolo\v{s} M., (2016), ``Acceleration of the charged particles due to chaotic scattering in the combined black hole gravitational field and asymptotically uniform magnetic field'', \textit{The European Physical Journal C}, 76, id.\ 32 (21 pp.)

\bibitem{Stu-etal:2020:Universe:}
Stuchl\'{\i}k Z.,  Kolo\v{s} M., Kov\'a\v{r} J., et al.\ (2020), ``Influence of cosmic repulsion and magnetic fields on accretion disks rotating around Kerr black holes'', \textit{Universe}, 6, 26

\bibitem{Stu-Kol-Tur:2021:Universe:}
Stuchl\'{\i}k Z.,  Kolo\v{s} M., Tursunov A. (2021), ``Penrose process: its variants and astrophysical applications'', \textit{Universe}, 7, 416

\bibitem{Stu-Kol-Tur:2022:PASJ:}
Stuchl\'{\i}k Z.,  Kolo\v{s} M., Tursunov A. (2022), ``Large-scale magnetic fields enabling fitting of the high-frequency QPOs observed around supermassive black holes'', \textit{Publications of the Astronomical Society of Japan}, 74, 1220

\bibitem{Tur-etal:2020:ApJ:}
Tursunov A., Stuchl\'{\i}k Z., Kolo\v{s} M. (2020), ``Supermassive black holes as possible sources of ultrahigh-energy cosmic rays'', \textit{The Astrophysical Journal}, 895, id.\ 14 (11 pp.)

\bibitem{tur20}
Tursunov A., Zaja{\v{c}}ek M., Eckart A., et al. (2020) ``Effect of electromagnetic interaction on Galactic center flare components'', \textit{The Astrophysical Journal}, 897, id.\ 99 (22 pp.)

\bibitem{wal72}
Wald R.~M. (1972), ``On uniqueness of the Kerr-Newman black holes'', \textit{J. Math. Phys.}, 13, 490

\bibitem{wal84}
Wald R.~M. (1984), General Relativity (Chicago: University of Chicago Press)

\end{thebibliography}


\eject
\section*{Short Biography of Authors}
{\textbf{Vladim\'{\i}r Karas} (*1960) is profesor of Theoretical physics and astrophysics at the Astronomical Institute of the Czech Academy of Sciences in Prague, where he has established the Relativistic Astrophysics program. Previously, Vladim\'{\i}r held a position of Associate Professor of astrophysics at Charles University and Visiting Scientist at various institutes abroad: Trieste, Rome, Paris, Warsaw, Baltimore, and Cambridge. He has developed fast numerical approaches to study effects of strong gravity via light variations from sources near a black hole. Vladim\'{\i}r is also interested in the motion of stars embedded in a dense nuclear star cluster, in particular, an interaction between stars and their environment in the immediate vicinity of galactic nuclei. More recent work deals with signatures of frame-dragging in twisted shapes of magnetic field lines: an interplay between Mach and Meissner effects near compact magnetic stars. Vladim\'{\i}r Karas served as the director of the Astronomical Institute in the period 2012-2022.}

{\textbf{Zden\v{e}k Stuchl\'{\i}k} (*1950) is profesor of Theoretical physics and astrophysics at the Institute of Physics of the Silesian University in Opava. His areas of professional interest include the structure of spacetimes around black holes and neutron stars and the propagation of light and the motion of test particles and perfect fluids in their vicinity. He has long been interested in the idea of using X-ray data from accreting compact objects to determine the basic physical parameters of the Universe, such as the cosmological constant or the tidal charge in the case of brane models, and its fundamental principles (the principle of Cosmic Censorship). Zden\v{e}k has collaborated with colleagues in Trieste, Rome, Oxford, Warsaw, and elsewhere. Zden\v{e}k Stuchl\'{\i}k is one of the founders of the Silesian University in Opava. Since 2020 he is the managing director of its Institute of Physics.}


\PublishersNote{}
\end{adjustwidth}
\end{document}